\documentclass{article}
\usepackage{amsmath}
\usepackage{amsfonts}
\usepackage{amssymb}
\usepackage{graphicx}
\providecommand{\U}[1]{\protect\rule{.1in}{.1in}}
\newtheorem{theorem}{Theorem}

\newtheorem{notation}[theorem]{Notation}

\newtheorem{definition}[theorem]{Definition}

\newtheorem{problem}[theorem]{Problem}

\newtheorem{remark}[theorem]{Remark}

\usepackage{epsfig}
\usepackage[all]{xy}
\usepackage{gensymb}
\usepackage{titlecaps}
\usepackage{verbatim}
\usepackage{float}
\usepackage{mathrsfs}  
\usepackage{bm}
\usepackage{multirow}
\usepackage{color,soul}

\DeclareSymbolFont{symbolsC}{U}{txsyc}{m}{n}
\SetSymbolFont{symbolsC}{bold}{U}{txsyc}{bx}{n}

\DeclareMathSymbol{\Diamonddot}{\mathord}{symbolsC}{144}
\DeclareMathSymbol{\Boxdot}{\mathbin}{symbols}{237}

\newcommand{\norm}[1]{\left\lVert#1\right\rVert}

\let\oldIEEEkeywords\IEEEkeywords
\def\IEEEkeywords{\oldIEEEkeywords\normalfont\bfseries\ignorespaces}

\begin{document}
	
\title{Robust Inference and Verification of Temporal Logic Classifier-in-the-loop Systems}

	\author{Zhe~Xu\thanks{Zhe~Xu is with the Oden Institute
		for Computational Engineering and Sciences, University of Texas,
		Austin, Austin, TX 78712, e-mail: zhexu@utexas.edu.}     
}

\maketitle

\section{Introduction}
Autonomous systems embedded with machine learning modules often rely on deep neural networks for classifying different objects of interest in the environment or different actions or strategies to take for the system. Due to the non-linearity and high-dimensionality of deep neural networks, the interpretability of the autonomous systems is compromised. Besides, the machine learning methods in autonomous systems are mostly data-intensive
and lack commonsense knowledge and reasoning that are natural to humans. In recent years, the use of formal methods have proven to improve both the interpretability and data-efficiency of autonomous systems \cite{xu2019joint}.

In this paper, we propose the framework of temporal logic classifier-in-the-loop systems. The temporal logic classifiers can output different actions to take for an autonomous system based on the environment, such that the behavior of the autonomous system can satisfy a given temporal logic specification. With the increasing development of artificial intelligence and machine learning, there has been a growing interest in learning (inferring) dense-time temporal logic formulas from system trajectories \cite{Kong2017,zheletter2,zhe2016,Bombara2016,zhe2015,zhe_ijcai2019,Asarin2012,Yan2019swarm,zhe2019ACCinfo,Hoxha2017,zheCDC2019GTL,Jin13}. Such temporal logic formulas have been used as high-level knowledge or specifications in many applications in robotics \cite{Allerton2019,zheACC2019DF,zhe_advisory,MH2019IFAC}, power systems \cite{zheACCstorageControl,zhe2017cascade,zhe_control,zheACC2018wind}, smart buildings \cite{zheCDCprivacy,zhe2019privacy}, agriculture \cite{cubuktepe2020policy}, etc. We use the temporal logic inference methods to infer temporal logic formulas that can classify different actions of the system based on the environment. Our approach is robust and provably-correct, as we can prove that the behavior of the autonomous system can satisfy a given temporal logic specification in the presence of (bounded) disturbances.

\section{Preliminaries}
\subsection{Metric Temporal Logic (MTL) for Discrete-time Trajectories}
	\label{MTL}   
	In this subsection, we briefly review the MTL that are interpreted over discrete-time trajectories~\cite{FainekosMTL}. Let $(\mathcal{X},d)$ be a metric space, where $x\in \mathcal{X}$ is a point, $d$ is a metric on $\mathcal{X}$. A set $\mathcal{AP}=\{\pi_1,\pi
	_2,\dots \pi_n\}$ is a set of atomic propositions, each of which can be
		either true or false.  The
		syntax of MTL is defined recursively as follows:
		\[
		\phi:=\top\mid \pi\mid\lnot\phi\mid\phi_{1}\wedge\phi_{2}\mid\phi_{1}\vee
		\phi_{2}\mid\phi_{1}\mathcal{U}_{\mathcal{I}}\phi_{2}\mid\phi_{1}\mathcal{S}_{\mathcal{I}}\phi_{2}
		\]
		where $\top$ stands for the Boolean constant True, $\pi$ is an atomic
		proposition, $\lnot$ (negation), $\wedge$(conjunction), $\vee$ (disjunction)
		are standard Boolean connectives, $\mathcal{U}$ is a temporal operator
		representing \textquotedblleft until\textquotedblright, $\mathcal{I}$ is a time interval of
		the form $I=[i_{1},i_{2})$. From \textquotedblleft
		until\textquotedblright($\mathcal{U}$), we
		can derive the temporal operators \textquotedblleft
		eventually\textquotedblright~$\Diamond_{\mathcal{I}}\phi=\top\mathcal{U}_{\mathcal{I}}\phi$ and
		\textquotedblleft always\textquotedblright~$\Box_{\mathcal{I}}\phi=\lnot\Diamond_{\mathcal{I}}\lnot\phi$. From \textquotedblleft
	since\textquotedblright($\mathcal{S}$), we can also derive the temporal operators \textquotedblleft
	eventually in the past\textquotedblright~$\Diamonddot_{\mathcal{I}}\phi=\top\mathcal{S}_{\mathcal{I}}\phi$ and
	\textquotedblleft always in the past\textquotedblright~$\boxdot_{\mathcal{I}}\phi=\lnot\Diamonddot_{\mathcal{I}}\lnot\phi$.
	We define a predicate mapping $\mathcal{O}:\mathcal{AP}\rightarrow2^\mathcal{X}$ such that for each $\pi \in \mathcal{AP}$ the corresponding set is $\mathcal{O}(\pi)\in \mathcal{X}$.\\
	
 A discrete-time trajectory $x=x(0), x(1),\dots, x(n_T)$ is a timed state sequence, with the corresponding time instants $t(0), t(1),\dots, t(n_T)$, where $n_T\in\mathbb{N}$, $x(k)\in \mathcal{X}$ and $t(k)\in \mathbb{R}_{\geqslant0}$ for every $k\in \{0,1,2,\dots,n_T\}$. We define $\Sigma_\mathcal{X}$ to be the set of all possible timed state sequences in the metric space $(\mathcal{X},d)$. We denote formula satisfiability using a membership function $\langle\langle\phi\rangle\rangle:=\Sigma_\mathcal{X}\times\mathbb{N}\rightarrow\mathbb{B}$ so that a discrete-time trajectory $x$ satisfies the formula $\phi$ at time instant $k$ when $\langle\langle\phi\rangle\rangle(x,k)=\top$. Then the Boolean semantics of MTL are defined recursively as follows:
	\[
	\begin{split}
	\langle\langle\top\rangle\rangle(x,k) :=& \top,\\
	\langle\langle \pi\rangle\rangle(x,k) :=& x(k)\in\mathcal{O}(\pi),\\
	\langle\langle \neg\phi\rangle\rangle(x,k) :=&\neg\langle\langle \phi\rangle\rangle(x,k),\\
	\langle\langle\phi_1\vee\phi_2\rangle\rangle(x,k):=&\langle\langle\phi_1\rangle\rangle(x,k)\vee\langle\langle\phi_2\rangle\rangle(x,k),\\
	\langle\langle\phi_1\mathcal{U}_{\mathcal{I}}\phi_{2}\rangle\rangle(x,k):=&\bigvee_{t(k')\in (t(k)+\mathcal{I})}\big(\langle\langle \phi_2\rangle\rangle(x,k')\wedge\bigwedge_{t(k)\le t(k'')<t(k')}\langle\langle \phi_1\rangle\rangle\\
	&(x,k'')\big), \\
	\langle\langle\phi_1\mathcal{S}_{\mathcal{I}}\phi_{2}\rangle\rangle(x,k):=&\bigvee_{t(k')\in (t(k)-\mathcal{I})}\big(\langle\langle \phi_2\rangle\rangle(x,k')\wedge\bigwedge_{t(k')\le t(k'')<t(k)}\langle\langle \phi_1\rangle\rangle\\
	&(x,k'')\big), 	
	\end{split}
	\]
where $t[k]+\mathcal{I}=\{t[k]+\tilde{t}\vert\tilde{t}\in\mathcal{I}\}$, $t[k]-\mathcal{I}=\{t[k]-\tilde{t}\vert \tilde{t}\in\mathcal{I}\}$.

We denote the distance from $x$ to a set $S$ as \textbf{dist}$_d(x,S)\triangleq$inf$\{d(x, y)\vert y\in cl(S)\}$ where $cl(S)$ denotes the closure of the set $S$, the depth of $x$ in $S$ as \textbf{depth}$_d(x,S)\triangleq$ \textbf{dist}$_d(x,\mathcal{X}\setminus S)$, the signed distance from $x$ to $S$ as
	\begin{equation}
	\textbf{Dist$_d(x,S)\triangleq$}%
	\begin{cases}
	-\textbf{dist}_d(x,S)& \mbox{if $x$ $\not\in \mathcal{X}$}\\  
	\textbf{depth}_d(x,S) & \mbox{if $x$ $\in \mathcal{X}$}
	\end{cases}                        
	\end{equation}
	
We use $\left[\left[\phi\right]\right](x, k)$ to denote the robustness estimate with which the discrete-time trajectory $x$ satisfies the specification $\phi$. The robust semantics of a formula $\phi$ with respect to $x$ are defined recursively as follows:
	\[
	\begin{split}
	\left[\left[\top\right]\right](x, k):=& +\infty,\\
	\left[\left[\pi\right]\right](x, k):=&\textbf{Dist$_d(x(k),\mathcal{O}(\pi))$},\\	
	\left[\left[\neg\phi\right]\right](x, k):=&-\left[\left[ \phi\right]\right](x, k),\\	
	\left[\left[\phi_1\wedge\phi_2\right]\right](x, k):=&\min\big(\left[\left[ \phi_1\right]\right](x, k),\left[\left[\phi_2\right]\right](x, k)\big),\\
	\left[\left[\phi_1\mathcal{U}_{\mathcal{I}}\phi_{2}\right]\right](x, k):=&\max_{t(k')\in (t(k)+\mathcal{I})}\Big(\min\big(\left[\left[\phi_2\right]\right](x, k'),\\& \min_{t(k)\le t(k'')<t(k')}\left[\left[\phi_1\right]\right]
	(x,k'')\big)\Big),\\
	\left[\left[\phi_1\mathcal{S}_{\mathcal{I}}\phi_{2}\right]\right](x, k):=&\max_{t(k')\in (t(k)-\mathcal{I})}\Big(\min\big(\left[\left[\phi_2\right]\right](x, k'),\\& \min_{t(k')\le t(k'')<t(k)}\left[\left[\phi_1\right]\right]
	(x,k'')\big)\Big).	 
	\end{split}
	\]

\subsection{One-clock Alternating Timed Automaton}
\label{OCATA_sec}
	
\begin{definition}
	\label{OCATA}
A one-clock alternating timed automaton (OCATA) \cite{Brihaye2013} is a tuple $\mathcal{A}=\{\mathcal{AP},\mathcal{L},\ell_0,\mathcal{F},\Delta\}$, where 
\begin{itemize}
\item $\mathcal{AP}$ is a set of atomic propositions;
\item $\mathcal{L}$ is a set of locations;
\item $\ell_0$ is the initial location;
\item $\mathcal{F}\subset\mathcal{L}$ is a set of accepting locations;
\item $\Delta:\mathcal{L}\times\mathcal{AP}\rightarrow\Gamma(\mathcal{L})$ is the transition function, where $\Gamma(\mathcal{L})$ denotes the set of formulas defined by
the following grammar:\\
\[
\gamma:=\top\mid\bot\mid\gamma_1\vee\gamma_2\mid\gamma_1\wedge\gamma_2\mid\ell
\mid c\bowtie g\mid c.\gamma
\]
where $g\in\mathbb{N}$, $\bowtie\in\{<,\le,>,\ge\}$, $\ell\in\mathcal{L}$, $c\bowtie g$ is a clock
constraint, $c.\gamma$ means that clock $c$ must be reset to 0.
\end{itemize}
\end{definition}

\section{Problem Formulation}
We consider the following discrete-time linear system:
\begin{align}
x(k+1)=Ax(k)+Bu(k),
\label{sys0}
\end{align}
where $x(\cdot)\in\mathcal{X}\subset\mathbb{R}^{n}$ is the state of the controlled agent, $A\in \mathbb{R}^{n \times n}$, $u(\cdot)\in\mathbb{R}^{m}$ is the control input that takes values from a discrete set $U=\{\bar{u}^1, \bar{u}^2, \dots, \bar{u}^M\}$.

\begin{notation}
	\label{note0}
We denote the solution of the linear system (\ref{sys0}) starting from $x(0)=x_0$ as $\xi(\cdot;x_0,u)$.
\end{notation}

We also denote $y$ as the discrete-time trajectory of the environment, $y(\cdot)\in\mathcal{Y}\subset\mathbb{R}^{p}$. We assume that the dynamical model of the environment is unknown.

\begin{problem}[feedforward controller design]
For the linear system (\ref{sys0}), design the control input $u(\cdot)$ from a discrete set $U=\{\bar{u}^1, \bar{u}^2, \dots, \bar{u}^M\}$ such that $\langle\langle\phi\rangle\rangle(q_{x,y}, 0)=\top$, where $\phi$ is a MTL specification of the system requirements, $q_{x,y}$ is a discrete-time feature trajectory generated through a feature map $Q:\mathcal{X}\times\mathcal{Y}\rightarrow \Sigma_{x,y}$, where $\Sigma_{x,y}$ is the feature space of both $x$ and $y$.
\label{problem_1}
\end{problem}

For example, the MTL specification $\phi$ could be ``the controlled robot should reach the shelf within 20 seconds while avoiding a moving obstacle''. When the dynamical model of the environment (e.g. the moving obstacle) is unknown and $q_{x,y}$ explicitly depends on both $x$ and $y$, the above problem generally cannot be solved using a pure model-based approach. However, a data-driven approach such as learning by human demonstrations could be useful. We assume that through human demonstrations, a set $J=\{(\xi^1(\cdot;x^1_0,u^1),y^1),$ $(\xi^2(\cdot;x^2_0,u^2),y^2), \dots, (\xi^N(\cdot;x^N_0,u^N),y^N)\}$ of finitely-many pairs of \textit{nominal} trajectories can be generated such that $\langle\langle\phi\rangle\rangle(q_{x^i,y^i}, 0)=\top$, $1\le i\le N$, where $\xi^i(\cdot;x^i_0,u^i)$ and $y^i$ are the $i$-th trajectory of the controlled agent and the environment, respectively, $q_{x^i,y^i}(k)=Q(\xi^i(k;x^i_0,u^i),y^i(k)),\forall k$. \\

In the following, we use the metric $d(a,b)=\norm{a-b}_M=(a-b)^TM(a-b)$ ($M$ is a matrix) for vectors $a$, $b$ and the metric $d_O(x,y)=\max\limits_{k}d(x(k), y(k))$ for discrete-time trajectories $x$, $y$. 
We denote $B(x_0, r)\triangleq\{x_0' \vert d(x_0,x_0')\le r\}$.\\

We intend to construct an MTL classifier, consisting of MTL formulas inferred from the past trajectory of the environment in the human demonstrations as decision criteria for selecting different inputs at each time instant. For example, if the traffic light is green and the front car with the speed less than the speed limit is moving away from the controlled car for the past 2 seconds, then the controlled car should select an input towards accelerating the speed. We call the inferred MTL formulas about the past trajectory of the environment as the \textit{environment decision MTL formulas}.

\begin{notation}
 \label{decision}	
We use $\psi$ to denote the environment decision MTL formula.
\end{notation}

To ensure causality of the MTL classifier, we only use the temporal operators \textquotedblleft
eventually in the past\textquotedblright~$\Diamonddot_{\mathcal{I}}$ and
\textquotedblleft always in the past\textquotedblright~$\boxdot_{\mathcal{I}}$ for the environment decision logic formulas.

For an environment decision MTL formula $\psi$ with temporal operators $\Diamonddot_{\mathcal{I}}$ or $\boxdot_{\mathcal{I}}$, the necessary length $\norm{\psi}$ is defined recursively as follows: 
\[                                                                   
\begin{split}
&\norm{\pi} =0, ~\norm{\lnot\psi} =\norm{\psi},\\
&\norm{\psi_{1}\wedge\psi_{2}}=\max(\norm{\psi_{1}},\norm{\psi_{2}}),\\
&\norm{\Diamonddot_{[t_1,t_2)}\psi}=\norm{\boxdot_{[t_1,t_2)}\psi}=\norm{\psi}+t_2.
\end{split}
\]

We use $h_{y}$ to denote the discrete-time feature trajectory of the environment generated through a feature map $H:\mathcal{Y}\rightarrow \Sigma_{y}$, where $\Sigma_{y}$ is the feature space of $y$. We assume that at every time instant $t(k)$, the discrete-time trajectory of the environment with past $D$ steps (with $t(k)-t(k-D)\ge\norm{\psi}$) is available. Note that here we allow negative time instants when $k<D$ to guarantee that $\psi$ can be always evaluated at any time instant $t(k)$.

Under the same or similar environment conditions, the ``correct'' input of the controlled agent could be totally different depending on what the controlled agent's task at the moment is. For example, with the MTL specification $\phi=\Diamond_{[0,10]}(p_{region 1}\wedge\Diamond_{[0,15]}p_{region 2})\wedge\Box_{[0,40]}\lnot p_{obstacle}$, which means ``the controlled agent should eventually reach region 1 within the next 10 time units and then eventually reach region 2 within the next 15 time units while avoiding a moving obstacle'', when the moving obstacle is far away from the controlled agent and the two regions, the ``correct'' input should be selected towards going to region 1 before region 1 is reached, and selected towards going to region 2 after region 1 is reached. Therefore, we first translate the MTL specification $\phi$ into a One-clock Alternating Timed Automaton (OCATA, see Definition \ref{OCATA} in Sec. \ref{OCATA_sec}) $\mathcal{A}_\phi$ with $\vert\mathcal{L}\vert$ different locations, accepting all and only the discrete-time trajectories satisfying formula $\phi$. We assume that $L\le\vert\mathcal{L}\vert$ locations are covered by the set $J$ through human demonstrations. Then we infer the environment decision MTL formulas in the $L$ different locations. \\
%
%
%
%
%
%
%
%

 
\begin{problem}[Robust MTL Classifier Inference]
Assume that $J=\{(\xi^1(\cdot;x^1_0,u^1),y^1),$ $(\xi^2(\cdot;x^2_0,u^2),y^2), \dots, (\xi^N(\cdot;x^N_0,u^N),y^N)\}$ is a set of $N$ pairs of nominal trajectories of system (\ref{sys0}), where $\langle\langle\phi\rangle\rangle(q_{x^i,y^i}, 0)=\top$, $1\le i\le N$, $q_{x^i,y^i}(k)=Q(\xi^i(k;x^i_0,u^i),y^i(k))$, $\forall k$. Given a positive number $\epsilon$, we seek to find $L$ sets of environment decision MTL formulas $\{\psi^{1}_{\ell},\dots,\psi^{P_\ell}_{\ell}\}$ ($1\le\ell\le L$, $1\le P_\ell\le M$ for each $\ell$), corresponding to $L$ sets of inputs $\hat{U}_{\ell}=\{\hat{u}^1_{\ell},\dots,\hat{u}^{P_\ell}_{\ell}\}\subset U=\{\bar{u}^1, \bar{u}^2, \dots, \bar{u}^M\}$, and the maximal positive numbers $\delta_c$, $\delta_e$ such that the followings are true:\\
(1) \textbf{$(\delta_c,\delta_e)$-robustness for system requirement $\phi$}: for every $i$, for any $\tilde{x}^i\triangleq\xi^i(\cdot;\tilde{x}^i_0,u^i)$ and $\tilde{y}^i$ such that $d(x^i_0, \tilde{x}^i_0)\le\delta_c$, $d_O(h_{y^i}, h_{\tilde{y}^i})\le\delta_e$, we have $\langle\langle\phi\rangle\rangle(q_{\tilde{x}^i,\tilde{y}^i}, 0)=\top$, where $q_{\tilde{x}^i,\tilde{y}^i}(k)=Q(\xi^i(k;\tilde{x}^i_0,u^i),\tilde{y}^i(k))$, $\forall k$.\\
(2) \textbf{$(\delta_c,\delta_e,\epsilon)$-robustness for soundness of MTL classifier}: for every $i$, $j$, $k$, and for any $\tilde{y}^i$ such that $d_O(h_{y^i}, h_{\tilde{y}^i})\le\delta_e$, if the current location of the OCATA $\mathcal{A}_\phi$ is $\ell$ and $\langle\langle\psi^j_{\ell}\rangle\rangle(h_{\tilde{y}^i}, k)=\top$, then $d(u^i(k),\hat{u}^j_{\ell})\le\epsilon$.\\
(3) \textbf{$(\delta_c,\delta_e)$-robustness for coverage of MTL classifier}: for every $i$, $k$, and for any $\tilde{y}^i$ such that $d_O(h_{y^i}, h_{\tilde{y}^i})\le\delta_e$, there exists $j$ such that $\langle\langle\psi^{j}_{\ell}\rangle\rangle(h_{\tilde{y}^i}, k)=\top$, with $\ell$ being the current location of the OCATA $\mathcal{A}_\phi$;\\
(4) \textbf{mutual exclusivity}: for every $\ell$ and every $j$, $\tilde{j}$ $(j\neq\tilde{j})$, at any time instant $k$, $\mathcal{B}_k(\psi^{j}_{\ell})\cap\mathcal{B}_k(\psi^{\tilde{j}}_{\ell})=\emptyset$, where $\mathcal{B}_k(\psi)$ denotes the set of all possible past trajectories of the environment that satisfies $\psi$ at time instant $k$. 
 \label{problem_2}
\end{problem}  

\begin{remark}
At each location $\ell$ of the OCATA $\mathcal{A}_\phi$, we infer $P_\ell$ $(P_\ell\le M)$ environment decision MTL formulas instead of exactly $M$ environment decision MTL formulas, to account for the occasions that under the same or similar environment in the same location $\ell$, the ``correct'' input may still not be unique. However, we assume that under such circumstances these ``correct'' inputs are similar to each other (with the maximal difference bounded by $\epsilon$).                   
\end{remark}

Intuitively, $(\delta_c,\delta_e)$-robustness for system requirement $\phi$ is to ensure that when we perturb the initial state of the controlled agent and perturb the environment trajectory by $\delta_c$ and $\delta_e$ respectively, while applying the same input signal with the nominal trajectory, the resulting trajectory $q_{\tilde{x}^i,\tilde{y}^i}$ still satisfies the MTL specification $\phi$ of system requirements; $(\delta_c,\delta_e,\epsilon)$-robustness for soundness of MTL classifier is to ensure that when we perturb each environment trajectory (with index $i$) by $\delta_e$, when the current location of the OCATA $\mathcal{A}_\phi$ is $\ell$ and the environment decision logic formula $\psi^j_{\ell}$ is satisfied by the past trajectories of the environment at time $t(k)$, the selected input $\hat{u}^j_{\ell}$ should be within $\epsilon$ distance from the original input $u^i(k)$; $(\delta_c,\delta_e)$-robustness for coverage of MTL classifier is to ensure that with the same perturbations of the environment, at any time instant there always exists an environment decision logic formula that is satisfied by the past trajectory of the environment; the mutual exclusivity is to make different environment decision MTL formulas mutually exclusive in each of the $L$ locations of the OCATA $\mathcal{A}_\phi$.   

 \begin{figure}[th]
 	\centering
 	\includegraphics[scale=0.6]{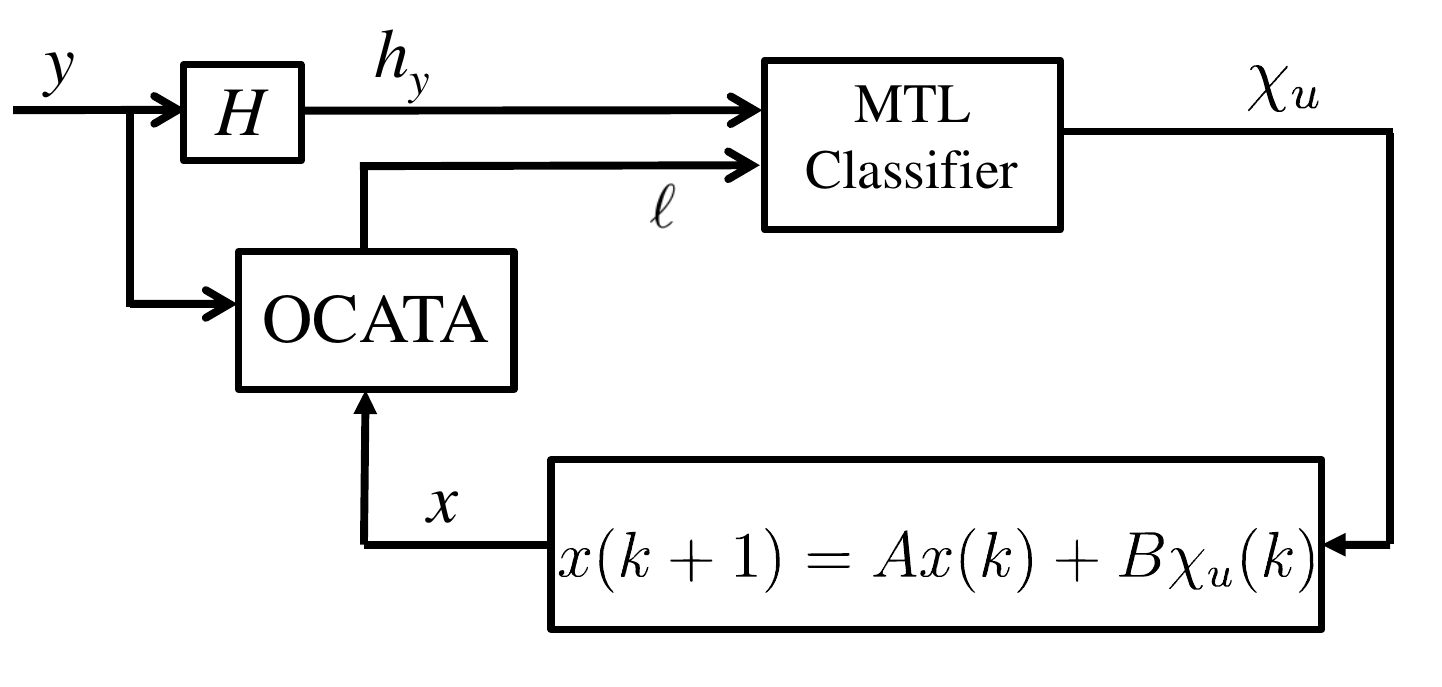}                                                                                                           
 	\caption{MTL Classifier-in-the-loop System $\mathcal{S}_{\rm{loop}}$.} \label{TL}
 \end{figure}

From the inferred MTL classifier, we can construct the following discrete-time MTL classifier-in-the-loop system $\mathcal{S}_{\rm{loop}}$:\\
\[
x(k+1)=Ax(k)+B\chi_u(k),
\] 
where
\begin{align}                                                 
\chi_u(k)=
\begin{cases}                                                                                                        
 \hat{u}^{1}_{\ell},
 	 ~~~~\mbox{if $\langle\langle\psi^{1}_{\ell}\rangle\rangle(h_{y}, k)=\top$},\\
 \hat{u}^{2}_{\ell}, ~~~~\mbox{if $\langle\langle\psi^{2}_{\ell}\rangle\rangle(h_{y}, k)=\top$},\\
 	~~~~~~~~~\vdots~~~~~~~~~~~~~~~~~~~~~~~~\vdots\\
 \hat{u}^{P_\ell}_{\ell}, ~~~\mbox{if $\langle\langle\psi^{P_\ell}_{\ell}\rangle\rangle(h_{y}, k)=\top$},\\
 	\end{cases}
\end{align}
where $\ell$ is the location of the OCATA $\mathcal{A}_\phi$ at the time instant $t(k)$, $\psi^{j}_{\ell}$ and $\hat{u}^{j}_{\ell}$ $(1\le \ell\le L$, $1\le P_\ell\le M$ for each $\ell$) are the inferred environment decision MTL formulas and control inputs respectively by solving Problem \ref{problem_2}, $y$ is the discrete-time trajectory of the environment, $h_{y}$ is the discrete-time feature trajectory of the environment.  

\begin{notation}
	\label{note1}
We denote the solution of the discrete-time MTL classifier-in-the-loop system $\mathcal{S}_{\rm{loop}}$ starting from $x(0)=x_0$ as $\xi(\cdot;x_0, \chi_u)$.
\end{notation}

The block diagram of the discrete-time MTL classifier-in-the-loop system $\mathcal{S}_{\rm{loop}}$ is shown in Fig. \ref{TL}. 

\begin{problem}[Verification of the MTL classifier-in-the-loop system]
Given the setting of Problem \ref{problem_2}, for a compact set $\mathcal{X}_{\rm{init}}\subset\bigcup\limits_{i} B(x^i_0, \delta_c)$ and a set $\mathcal{Y}_{\rm{tube}}$ of infinite trajectories of the environment where for any $y\in\mathcal{Y}_{\rm{tube}}$ there exists $i$ such that $d_O(h_{y}, h_{y^i})\le\delta_e$, design a mechanism to verify whether $\langle\langle\phi\rangle\rangle(\tilde{q}_{x,y}, k)=\top$ holds 
for any $x_0\in\mathcal{X}_{\rm{init}}$ and any $y\in\mathcal{Y}_{\rm{tube}}$, where $\phi$ is the MTL specification of the system requirements, $\tilde{q}_{x,y}(k)=Q(\xi(k;x_0,\chi_u), y(k)), \forall k$.
\label{problem_3}
\end{problem}

\bibliographystyle{IEEEtran}
\bibliography{zheloopref}

\begin{thebibliography}{10}
\providecommand{\url}[1]{#1}
\csname url@samestyle\endcsname
\providecommand{\newblock}{\relax}
\providecommand{\bibinfo}[2]{#2}
\providecommand{\BIBentrySTDinterwordspacing}{\spaceskip=0pt\relax}
\providecommand{\BIBentryALTinterwordstretchfactor}{4}
\providecommand{\BIBentryALTinterwordspacing}{\spaceskip=\fontdimen2\font plus
\BIBentryALTinterwordstretchfactor\fontdimen3\font minus
  \fontdimen4\font\relax}
\providecommand{\BIBforeignlanguage}[2]{{%
\expandafter\ifx\csname l@#1\endcsname\relax
\typeout{** WARNING: IEEEtran.bst: No hyphenation pattern has been}%
\typeout{** loaded for the language `#1'. Using the pattern for}%
\typeout{** the default language instead.}%
\else
\language=\csname l@#1\endcsname
\fi
#2}}
\providecommand{\BIBdecl}{\relax}
\BIBdecl

\bibitem{xu2019joint}
Z.~Xu, I.~Gavran, Y.~Ahmad, R.~Majumdar, D.~Neider, U.~Topcu, and B.~Wu,
  ``Joint inference of reward machines and policies for reinforcement
  learning,'' in \emph{Proc. International Conference on Automated Planning and
  Scheduling (ICAPS), Special Track on Planning and Learning}, 2020.

\bibitem{Kong2017}
Z.~Kong, A.~Jones, and C.~Belta, ``Temporal logics for learning and detection
  of anomalous behavior,'' \emph{IEEE Trans. Automatic Control}, vol.~62,
  no.~3, pp. 1210--1222, March 2017.

\bibitem{zheletter2}
Z.~Xu, M.~Birtwistle, C.~Belta, and A.~Julius, ``A temporal logic inference
  approach for model discrimination,'' \emph{IEEE Life Sciences Letters},
  vol.~2, no.~3, pp. 19--22, Sept 2016.

\bibitem{zhe2016}
\BIBentryALTinterwordspacing
Z.~Xu and A.~A. Julius, ``Census signal temporal logic inference for multiagent
  group behavior analysis,'' \emph{IEEE Trans. Autom. Sci. and Eng.}, 2016, in
  press. [Online]. Available:
  \url{http://ieeexplore.ieee.org/document/7587357/}
\BIBentrySTDinterwordspacing

\bibitem{Bombara2016}
\BIBentryALTinterwordspacing
G.~Bombara, C.-I. Vasile, F.~Penedo, H.~Yasuoka, and C.~Belta, ``A decision
  tree approach to data classification using signal temporal logic,'' in
  \emph{Proceedings of the 19th International Conference on Hybrid Systems:
  Computation and Control}, ser. HSCC '16.\hskip 1em plus 0.5em minus
  0.4em\relax New York, NY, USA: ACM, 2016, pp. 1--10. [Online]. Available:
  \url{http://doi.acm.org/10.1145/2883817.2883843}
\BIBentrySTDinterwordspacing

\bibitem{zhe2015}
Z.~Xu, C.~Belta, and A.~Julius, ``Temporal logic inference with prior
  information: An application to robot arm movements,'' \emph{IFAC Conference
  on Analysis and Design of Hybrid Systems (ADHS)}, pp. 141 -- 146, 2015.

\bibitem{zhe_ijcai2019}
\BIBentryALTinterwordspacing
Z.~Xu and U.~Topcu, ``Transfer of temporal logic formulas in reinforcement
  learning,'' in \emph{Proc. IJCAI'2019}, 7 2019, pp. 4010--4018. [Online].
  Available: \url{https://doi.org/10.24963/ijcai.2019/557}
\BIBentrySTDinterwordspacing

\bibitem{Asarin2012}
E.~Asarin, A.~Donz{\'e}, O.~Maler, and D.~Nickovic, ``Parametric identification
  of temporal properties,'' in \emph{Proc. Second Int. Conf. Runtime
  Verification}, Berlin, Heidelberg, 2012, pp. 147--160.

\bibitem{Yan2019swarm}
R.~{Yan}, Z.~{Xu}, and A.~{Julius}, ``Swarm signal temporal logic inference for
  swarm behavior analysis,'' \emph{IEEE Robotics and Automation Letters},
  vol.~4, no.~3, pp. 3021--3028, 2019.

\bibitem{zhe2019ACCinfo}
Z.~{Xu}, M.~{Ornik}, A.~A. {Julius}, and U.~{Topcu}, ``Information-guided
  temporal logic inference with prior knowledge,'' in \emph{2019 American
  Control Conference (ACC)}, July 2019, pp. 1891--1897.

\bibitem{Hoxha2017}
\BIBentryALTinterwordspacing
B.~Hoxha, A.~Dokhanchi, and G.~Fainekos, ``Mining parametric temporal logic
  properties in model-based design for cyber-physical systems,''
  \emph{International Journal on Software Tools for Technology Transfer}, Feb
  2017. [Online]. Available: \url{http://dx.doi.org/10.1007/s10009-017-0447-4}
\BIBentrySTDinterwordspacing

\bibitem{zheCDC2019GTL}
Z.~{Xu}, A.~J. {Nettekoven}, A.~{Agung Julius}, and U.~{Topcu}, ``Graph
  temporal logic inference for classification and identification,'' in
  \emph{2019 IEEE 58th Conference on Decision and Control (CDC)}, Dec 2019, pp.
  4761--4768.

\bibitem{Jin13}
X.~Jin, A.~Donze, J.~V. Deshmukh, and S.~A. Seshia, ``Mining requirements from
  closed-loop control models,'' in \emph{Proc. Int. Conf. Hybrid Systems:
  Computation and Control}, 2013, pp. 43--52.

\bibitem{Allerton2019}
Z.~{Xu}, F.~M. {Zegers}, B.~{Wu}, W.~{Dixon}, and U.~{Topcu}, ``Controller
  synthesis for multi-agent systems with intermittent communication. a metric
  temporal logic approach,'' in \emph{Allerton'19}, pp. 1015--1022.

\bibitem{zheACC2019DF}
Z.~Xu, K.~Yazdani, M.~T. Hale, and U.~Topcu, ``Differentially private
  controller synthesis with metric temporal logic specifications,'' in \emph{To
  appear in Proc. International Conference on Autonomous Agents and Multiagent
  Systems (AAMAS)}, 2020.

\bibitem{zhe_advisory}
Z.~Xu, S.~Saha, B.~Hu, S.~Mishra, and A.~A. Julius, ``Advisory temporal logic
  inference and controller design for semiautonomous robots,'' \emph{IEEE
  Trans. Autom. Sci. Eng.}, pp. 1--19, 2018.

\bibitem{MH2019IFAC}
M.~Hibbard, Y.~Savas, Z.~{Xu}, A.~A. {Julius}, and U.~{Topcu}, ``Minimizing the
  information leakage of high-level task specifications,'' in \emph{21st IFAC
  World Congress}, 2020.

\bibitem{zheACCstorageControl}
Z.~Xu, A.~Julius, and J.~H. Chow, ``Optimal energy storage control for
  frequency regulation under temporal logic specifications,'' in \emph{2017
  American Control Conference (ACC)}, May 2017, pp. 1874--1879.

\bibitem{zhe2017cascade}
Z.~Xu, A.~A. Julius, and J.~H. Chow, ``Robust testing of cascading failure
  mitigations based on power dispatch and quick-start storage,'' \emph{IEEE
  Systems Journal}, vol.~PP, no.~99, pp. 1--12, 2017.

\bibitem{zhe_control}
Z.~Xu, A.~Julius, and J.~H. Chow, ``Energy storage controller synthesis for
  power systems with temporal logic specifications,'' \emph{IEEE Systems
  Journal}, Early access on IEEE Xplore.

\bibitem{zheACC2018wind}
Z.~Xu, A.~A. Julius, and J.~H. Chow, ``Coordinated control of wind turbine
  generator and energy storage system for frequency regulation under temporal
  logic specifications,'' in \emph{Proc. Amer. Control Conf.}, 2018, pp.
  1580--1585.

\bibitem{zheCDCprivacy}
Z.~Xu, S.~Saha, and A.~Julius, ``Provably correct design of observations for
  fault detection with privacy preservation,'' in \emph{IEEE Conference on
  Decision and Control (CDC), Melbourne, Australia, 2017}.

\bibitem{zhe2019privacy}
Z.~Xu and A.~A. Julius, ``Robust temporal logic inference for provably correct
  fault detection and privacy preservation of switched systems,'' \emph{IEEE
  Systems Journal}, vol.~13, no.~3, pp. 3010--3021, 2019.

\bibitem{cubuktepe2020policy}
M.~Cubuktepe, Z.~Xu, and U.~Topcu, ``Policy synthesis for factored mdps with
  graph temporal logic specifications,'' in \emph{Proc. International
  Conference on Autonomous Agents and Multiagent Systems (AAMAS)}, 2020.

\bibitem{FainekosMTL}
G.~E. Fainekos and G.~J. Pappas\vspace{0mm}, ``Robustness of temporal logic
  specifications,'' in \emph{Formal Approaches to Testing and Runtime
  Verification, in: LNCS, vol. 4262, Springer, 2006}.

\bibitem{Brihaye2013}
\BIBentryALTinterwordspacing
T.~Brihaye, M.~Esti{\'{e}}venart, and G.~Geeraerts, ``On {MITL} and alternating
  timed automata,'' \emph{CoRR}, vol. abs/1304.2814, 2013. [Online]. Available:
  \url{http://arxiv.org/abs/1304.2814}
\BIBentrySTDinterwordspacing

\end{thebibliography}

\end{document}